\documentclass{article}
\usepackage{authblk, amsmath,graphicx}
\usepackage{arxiv}
\usepackage{amssymb }
\usepackage{algorithm}
\usepackage[noend]{algpseudocode}
\usepackage{cite, appendix}

\DeclareMathOperator*{\argmin}{arg\,min}
\def\BState{\State\hskip-\ALG@thistlm}
\newcommand{\pdiff}[1]{\frac{\partial}{\partial {#1}}}
\newcommand{\gcon}{\frac{2 \pi i}{\lambda}}
\newcommand{\gl}{\textbf{g}_\ell}
\newcommand{\Tl}{\mtx{T}_\ell}
\newcommand{\cgl}{\overline{\textbf{g}_\ell}}
\newcommand{\cx}{\bar{\vct{x}}}
\newcommand{\jac}{\mtx{J}}
\newcommand{\cplxvct}[1]{
	\begin{bmatrix}
		{#1}      \\
		\\
		\overline{{#1}}       \\
\end{bmatrix}}
\newcommand{\iterdiff}{\vct{x}_{\tau+1}- \vct{x}_{\tau}}
\usepackage{hyperref,graphicx,amsmath,amsfonts,amssymb,bm,url,breakurl,epsfig,epsf,color,MnSymbol,mathbbol,fmtcount,semtrans,caption,subcaption,multirow,comment, boldline}
\usepackage[noend]{algpseudocode}
\usepackage{wrapfig}
\usepackage{amssymb}

\usepackage[utf8]{inputenc} 
\usepackage[T1]{fontenc}    
\usepackage{url}            
\usepackage{booktabs}       
\usepackage{amsfonts}       
\usepackage{nicefrac}       
\usepackage{microtype}      

\makeatletter
\providecommand*{\boxast}{%
  \mathbin{
    \mathpalette\@boxit{*}%
  }%
}
\newcommand*{\@boxit}[2]{%
  \sbox0{$\m@th#1\Box$}%
  \ifx#1\displaystyle \ht0=\dimexpr\ht0+.05ex\relax \fi
  \ifx#1\textstyle \ht0=\dimexpr\ht0+.05ex\relax \fi
  \ifx#1\scriptstyle \ht0=\dimexpr\ht0+.04ex\relax \fi
  \ifx#1\scriptscriptstyle \ht0=\dimexpr\ht0+.065ex\relax \fi
  \sbox2{$#1\vcenter{}$}
  \rlap{%
    \hbox to \wd0{%
      \hfill
      \raisebox{%
        \dimexpr.5\dimexpr\ht0+\dp0\relax-\ht2\relax
      }{$\m@th#1#2$}%
      \hfill
    }%
  }%
  \Box
}
\makeatother

  \makeatletter
\def\BState{\State\hskip-\ALG@thistlm}
\makeatother

  \usepackage{mathtools}

\usepackage{titlesec}

\usepackage{tikz}
\usepackage{pgfplots}
\usetikzlibrary{pgfplots.groupplots}

\titleformat{\paragraph}
{\normalfont\normalsize\bfseries}{\theparagraph}{1em}{}
\titlespacing*{\paragraph}
{0pt}{3.25ex plus 1ex minus .2ex}{1.5ex plus .2ex}

\usepackage{movie15}

\usepackage{caption}
\usepackage[bottom,hang,flushmargin]{footmisc} 

\newcommand{\tsn}[1]{{\left\vert\kern-0.25ex\left\vert\kern-0.25ex\left\vert #1 
    \right\vert\kern-0.25ex\right\vert\kern-0.25ex\right\vert}}

\definecolor{darkred}{RGB}{150,0,0}
\definecolor{darkgreen}{RGB}{0,150,0}
\definecolor{darkblue}{RGB}{0,0,200}
\hypersetup{colorlinks=true, linkcolor=darkred, citecolor=darkgreen, urlcolor=darkblue}

\newtheorem{theorem}{Theorem}[section]

\newcommand{\beq}{\begin{equation}}

\newcommand{\eeq}{\end{equation}}

\newcommand{\twonorm}[1]{\left\|#1\right\|_{\ell_2}}

\newcommand{\infnorm}[1]{\left\|#1\right\|_{\ell_\infty}}

\newcommand{\abs}[1]{\left|#1\right|}

\definecolor{emmanuel}{RGB}{255,127,0}

\newcommand{\vct}[1]{\bm{#1}}
\newcommand{\mtx}[1]{\bm{#1}}

\numberwithin{equation}{section} 

\def \endprf{\hfill {\vrule height6pt width6pt depth0pt}\medskip}

\title{3D Phase retrieval at nano-scale via Accelerated Wirtinger Flow}
\author{Zalan Fabian, Justin Haldar, Richard Leahy, and Mahdi Soltanolkotabi}
\affil{Department of Electrical and Computer Engineering, University of Southern California}
\begin{document}
	%
	\maketitle
	\begin{abstract}
		Imaging 3D nano-structures at very high resolution is crucial in a variety of scientific fields. However, due to fundamental limitations of light propagation we can only measure the object indirectly via 2D intensity measurements of the 3D specimen through highly nonlinear projection mappings where a variety of information (including phase) is lost.
		 Reconstruction therefore involves inverting highly non-linear and seemingly non-invertible mappings. 
		In this paper, we introduce a novel technique where the 3D object is directly reconstructed from an accurate non-linear propagation model. Furthermore, we characterize the ambiguities of this model and leverage a priori knowledge to mitigate their effect and also significantly reduce the required number of measurements and hence the acquisition time. We demonstrate the performance of our algorithm via numerical experiments aimed at nano-scale reconstruction of 3D integrated circuits. Moreover, we provide rigorous theoretical guarantees for convergence to stationarity.
	\end{abstract}
	
	Imaging nano-structures at fine resolution has become increasingly important in diverse fields of science and engineering. For instance, quality control/examination of modern multi-layered integrated circuits requires detailed imaging of intricate 3D structures with $10$nm features. Similarly, real-time non-destructive imaging of biological specimens, such as protein complexes, on the molecular scale could provide invaluable insight into many biological processes that are little understood. Imaging at finer resolution necessitates high-energy beams with shorter wavelengths. Building optical components such as mirrors and lenses on this scale is very difficult and often phaseless coherent diffraction methods are required. This necessity triggered a major revival in phaseless imaging techniques and experiments \cite{phaseless:abbey2008keyhole,phaseless:clark2013ultrafast, phaseless:miao1999extending, phaseless:nelson2010high, phaseless:pfeifer2006three, phaseless:shapiro2005biological,ptychoexp:deng2015simultaneous, ptychoexp:dierolf_nature, ptychoexp:holler2017high, ptychoexp:shapiro2014chemical, ptychoexp:thibault2008high, fourierptychoexp:horstmeyer2015digital, fourierptychoexp:tian20153d, fourierptychoexp:tian2015computational, fourierptychoexp:zheng2013wide,partcohptychoexp:chang2018partially} as well as algorithms for phase retrieval. See \cite{awf:xu2018accelerated} for a comprehensive overview of algorithmic approaches and \cite{wf:candes2015phase, pradditional:chandra2017phasepack,  pradditional:waldspurger2018phase} for theoretical work. Some authors leverage prior knowledge on the signal structure such as sparsity \cite{pradditional:jaganathan2012recovery, pradditional:soltanolkotabi2019structured} in order to further decrease the necessary number of measurements. 

Despite all of this recent progress on phaseless reconstruction methods, there has been significantly less focus on 3D imaging at nano-scale. We briefly discuss a few recent efforts.  \cite{ptycho3d:maiden2012multisclice} uses a multi-slice approach to image thick specimens in 3D, where the wave front is propagated through the object layer by layer. Authors in \cite{ptycho3d:tian20153d} use this multi-slice forward model combined with Fourier-ptychography to successfully reconstruct thick biological samples. \cite{lamino3d:myagotin2013efficient} uses filtered backprojection for the reconstruction of flat specimens. \cite{ptychoexp:dierolf_nature} uses a two step approach where they first reconstruct 2D projections of the object from phaseless measurements, then obtain the 3D structure via tomography from the 2D reconstructions.
A more recent line of work investigates a joint technique that alternates between a ptychography step on exit waves and a tomographic reconstruction step on the object based on the updated projections \cite{pct:gursoy2017direct, pct:aslan2019joint, pct:nikitin2019photon, pct:chang2019iterative}. Most of these techniques typically use a first-order approximation of the projections in the tomography step due to the challenges introduced by the non-linearity. Even though the linear regime provides a good estimation for small biological samples, it becomes increasingly inaccurate for extended specimens and for materials used in electronics. An additional challenge of 3D imaging at very fine resolution is the extremely sensitive calibration process that highly increases data acquisition time.
%

In this work, we introduce a 3D reconstruction technique where the object is reconstructed directly in lieu of separating reconstruction into ptychography and tomography steps or alternating between those two as in prior work. Furthermore, we use a highly non-linear wave propagation model without linear approximation. We expect this model to be more accurate than the linear approximation, especially for larger specimens where the path length of the beam passing through the object is longer. Our work builds upon AWF \cite{awf:xu2018accelerated}, an accelerated optimization technique used for 2D phase retrieval. We extend this framework to 3D reconstruction by directly incorporating tomography in the algorithm and  by adding weighted TV-regularization, which we term 3D Accelerated Wirtinger Flow (3D-AWF). We show that the merit of TV-regularization is threefold: (1) it offers a computationally inexpensive method to alleviate the effect of ambiguities introduced by the non-linear model by leveraging prior knowledge, (2) it significantly accelerates data acquisition by reducing the number of measurements needed for a given level of reconstruction accuracy and (3) effectively incorporates the structure of integrated circuits by promoting a piecewise constant reconstruction. We demonstrate through numerical simulations on realistic chip data that our non-linear model results in significantly more accurate reconstructions compared to its linear approximation. Moreover, we provide mathematically rigorous guarantees for convergence of our algorithm. 
	\section{Phaseless imaging in 3D} \label{sec:model}
We are interested in reconstructing the complex valued 3D refractive index of the object, where we model the object as shifts of a voxel basis function over a cubic lattice. 
Let $\vct{x} \in \mathbb{C}^N$ represent the complex refractive index of the discretized 3D object obtained from vectorizing $\mtx{X} \in \mathbb{C}^{N_1 \times N_2 \times N_3}$ for which $\mtx{X}_{n_1,n_2,n_3}$ is the complex refractive index at voxel $(n_1,n_2,n_3)$ on a cubic lattice. Here, $N_1, N_2$ and $N_3$ denote the number of voxels along each dimension and obey $N_1N_2N_3=N$. Here, the object is of the form $\vct{x} = \vct{d} + i \vct{b}$ with $\vct{d}, \vct{b} \in \mathbb{R}^N$ with $\vct{d}$ denoting the phase shift and $\vct{b}$ the attenuation associated with wave propagation through the object, and $i$ the imaginary unit. Our forward model consists of two stages. The first stage consists of applying a non-linear projection to the 3D object resulting in a 2D complex exit wave. Then, the exit wave is passed through a linear mapping and its magnitude is measured in the far field.

\subsection{From 3D object to 2D exit waves.}
\begin{figure}
	\centering
	\includegraphics[scale=0.99]{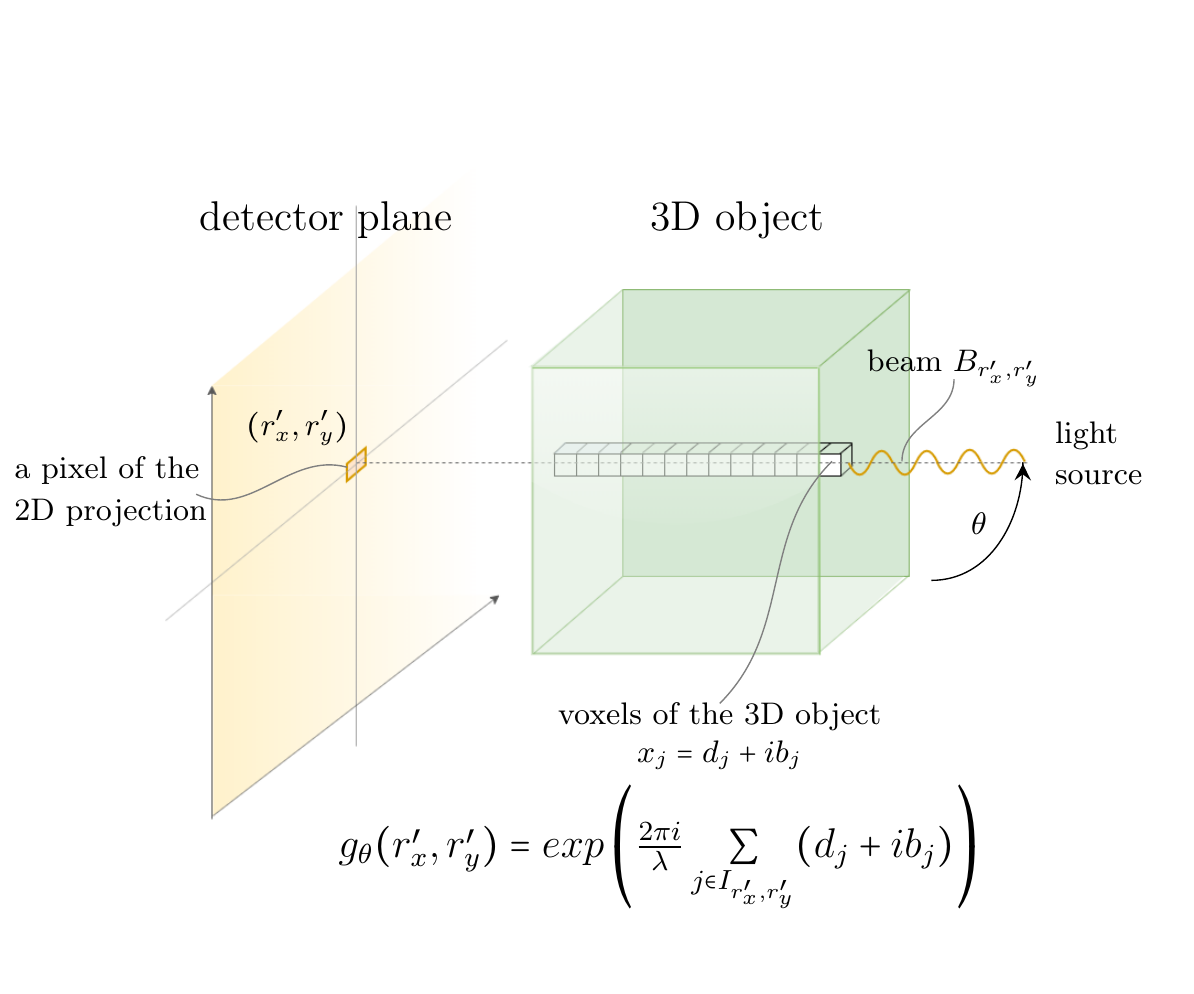}
	\caption{Projection approximation of wave propagation. Incident beam $B_{r_{x}', r_{y}'}$ passes through the 3D object, and produces a pixel in projection plane at $(r_{x}', r_{y}')$. $I_{r_{x}', r_{y}'}$ denotes the set of voxel indices intersected by the beam. \label{fig:proj}}
\end{figure}

Let $\Tl \in \mathbb{R}^{P\times N}$ represent the part of the conventional
Radon transform projection operator corresponding to the $\ell$th projection angle. Based on the projection approximation of wave propagation \cite{ptychoexp:dierolf_nature}(Fig. \ref{fig:proj}), for a wavelength $\lambda$ the mapping from $\vct{x}$ to
the discretized exit wave in orientation $\ell$ can be represented as
\begin{equation}\label{eq:gl}
	\gl := \gl(\vct{x})= exp\left(\gcon \Tl (\vct{d} + i \vct{b})\right),
	\vspace{-0.2cm}
\end{equation}
where exponentiation should be interpreted element-wise.

\subsection{From 2D exit waves to phaseless measurements.}
\begin{figure}
	\centering
	\includegraphics[scale=0.39]{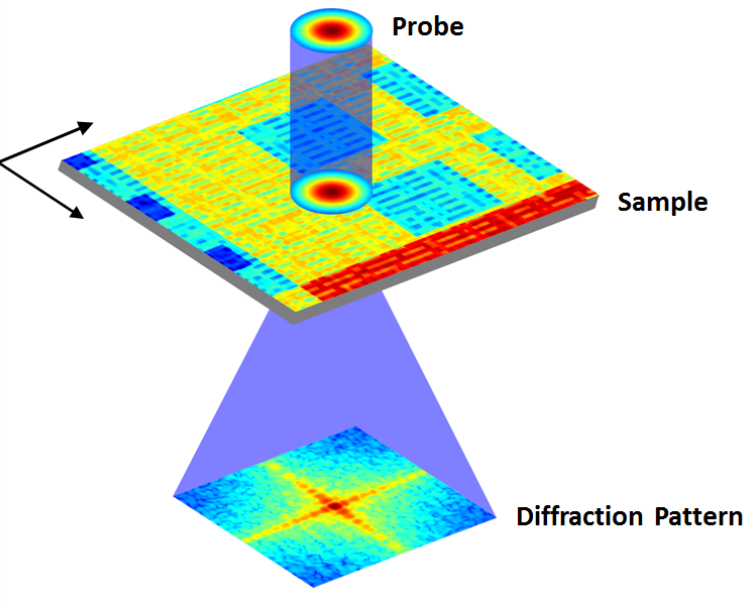}
	\caption{In ptychography, the sample is illuminated by a probe function from various angles. The diffraction pattern in the far field is the Fourier transform of the exit wave multiplied by the probe function. \label{fig:ptych}}
\end{figure}

In 3D ptychography, a sample is illuminated with several different illumination functions (or "probes") from $L$ different orientations and the corresponding diffraction patterns for each probe are measured by a detector in the far field (Fig. \ref{fig:ptych}). In many cases, the different probe functions $p_k(\vct{r}')$ are obtained as different spatial shifts of the same basic probe function. Let $g_\ell(\vct{r}')$ represent the 2D exit wave as a function of the spatial position in projection plane $\vct{r}' = (r_x', r_y')$ (\eqref{eq:gl} is the corresponding discretization). Then, the complex field at the detector plane resulting from the $k$th probe in orientation $\ell$ is given by
$
\zeta_{k,\ell}(\vct{r}') = \mathcal{F}\left\{p_k(\vct{r}') g_\ell(\vct{r}') \right\},
$
where $\mathcal{F}$ denotes the Fourier transform. However, we cannot sense the complex far field directly, only its magnitude. Therefore our phaseless measurement corresponding to the $k$th probe and $\ell$th illumination angle takes the form $\vct{y}_{k,\ell} = \abs{\zeta_{k,\ell}}$. All measurements obtained in the  $\ell$th orientation can be written in the more compact form $\vct{y}_\ell = \abs{\mtx{A} \gl}$, where $\mtx{A}$ represents the ptychographic propagation model described in \cite{awf:xu2018accelerated}.

\subsection{Ambiguity challenge.}
\textit{Tomography.} Recovering the phase of the ground truth object based on phaseless measurements is only possible up to some ambiguity factors. First, the mapping from the object $\vct{x}$ to the 2D projections $\{\mtx{T}_\ell\vct{x}\}_{\ell=1}^L$ may not invertible and therefore the 2D projection images may result from infinitely many possible 3D objects. This ambiguity is rather pronounced when we only have measurements from a few orientations. However, with a sufficiently large number of angles, the mapping is typically invertible. 

\textit{Global phase.} Another source of ambiguity arises from the phaseless measurements. Recovering the 2D exit waves is only possible up to a phase factor, since the magnitude measurements are invariant to a global shift in phase.

\textit{Phase wrapping.} Phase wrapping is another source of ambiguity that appears in the 2D exit waves and originates in the projection model in \eqref{eq:gl}. Specifically, let $\vct{x}^* = \vct{d}^* + i \vct{b}^*$ be the ground truth object we wish to reconstruct,  $\widehat{\vct{x}} = \widehat{\vct{d}} + i \widehat{\vct{b}}$ be the estimate obtained from our reconstruction algorithm, and $\widetilde{\vct{d}}:=\widehat{\vct{d}}-\vct{d}^*$ the error in the real part of the reconstruction. Then the reconstructed exit wave at orientation $\ell$ takes the form
\begin{align}
\label{ambg}
\widehat{\vct{g}}_\ell &= e^{-\frac{2\pi}{\lambda}\Tl\widehat{\vct{b}}} \cdot e^{i\frac{2\pi}{\lambda}\Tl \vct{d}^*} \cdot e^{i\frac{2\pi}{\lambda}\Tl \widetilde{\vct{d}}}.
\end{align}
From this identity it is clear that if $\widetilde{\vct{d}}$ is such that $\Tl \widetilde{\vct{d}} = \lambda \vct{k}_\ell$ for any $\vct{k}_\ell \in \mathbb{Z}^{N}$ will be consistent with the measurements and one can not hope to differentiate between the reconstruction and the ground truth exit wave. This effect translates to an ambiguity in the real part of each voxel of the 3D object: $\widehat{\vct{d}}$ and $\widehat{\vct{d}} + \lambda \vct{k}$ are indistinguishable for any $\vct{k} \in \mathbb{Z}^N$ in our model. 
To elaborate further, consider an incident beam $B^l_j$ at angle $l$ that produces pixel $\hat{r}^l_j$ on the projection image. Denote $I:\{i| x_i \in B^l_j\}$ the set of indices of voxels intersected by the beam. Explicitly writing out the Radon-transform for the real part of this pixel is simply a sum in the discrete case:
$$\Re( \hat{r}^\ell_j )=  \sum_{i \in I} \hat{d}_i=  \sum_{i \in I} d_i^* +\sum_{i \in I} \tilde{d}_i. $$
Assume that the voxel-wise reconstruction error can be written as 
$\tilde{d}_i = k_i \lambda, ~k_i \in \mathbb{Z} ~ \forall i,$
and in this case
$$\Re(\hat{r}^\ell_j )=  \sum_{i \in B} d_i^* +\lambda \sum_{i \in B} k_i =\sum_{i \in B} d_i^* +\lambda k',~~ k' \in \mathbb{Z} $$
which results in the same exit wave pixel as the ground truth object and therefore indistinguishable from the ground truth in our model. Moreover, due to phase wrapping we lose all information on each individual $k_i$. 



	\section{Reconstruction via 3D-AWF}
Our goal is to find $\vct{x} \in \mathbb{C}^N$ that best explains our phaseless measurements under the propagation model. Formally, we solve the optimization problem 
\begin{equation}\label{eq:optprob}
    \hat{\vct{x}} = \argmin_{\vct{x} \in \mathbb{C}^N}\quad\mathcal{L}(\vct{x}) + \lambda_{TV}\textbf{TV}_{3D}(\vct{x}; \vct{w} )= \argmin_{\vct{x} \in \mathbb{C}^N}\quad\mathcal{L}_{total}(\vct{x}),
\end{equation}
where $\lambda_{TV} \in [0, \infty )$ is the regularization strength and
$$
\mathcal{L}(\vct{x}) := \sum_{l=1}^{L} \|\vct{y}_{l}- |\mtx{A} \gl|\|_2^2.
$$
The second term penalizes the weighted total variation of the reconstruction defined as
$$
\textbf{TV}_{3D}(\vct{x}; \vct{w}) = \sum_{i,j,k} \left(w_x\abs{x_{i+1,j,k}-x_{i,j,k}}+w_y\abs{x_{i,j+1,k}-x_{i,j,k}}
+w_z\abs{x_{i,j,k+1}-x_{i,j,k}}\right),
$$
where $\vct{w} = [w_x, w_y, w_z]$ is a fixed vector of non-negative weights that can be used to leverage prior knowledge on the structure of the object along different spatial dimensions.
The optimization problem in \eqref{eq:optprob} is nonconvex and in general  does not admit a closed form solution. Classical gradient descent requires a differentiable loss landscape and the loss in \eqref{eq:optprob} is not complex differentiable. However, this does not pose a significant challenge since the loss function is differentiable except for isolated points, and we can define generalized gradients at non-differentiable points \cite{book:clarke1990optimization}. We use the notion of Wirtinger-derivatives and apply a proximal variant of AWF \cite{awf:xu2018accelerated}, which we call 3D-AWF with update rule
 \begin{align} \label{eq:awfprox}
\vct{z}_{\tau+1} &= \vct{x}_{\tau} + \beta_{\tau}(\vct{x}_{\tau} - \vct{x}_{\tau-1}) - \mu_{\tau} \nabla \mathcal{L}( \vct{x}_{\tau} + \beta_{\tau}(\vct{x}_{\tau} - \vct{x}_{\tau-1})) \nonumber\\
\vct{x}_{\tau+1} &= prox_{TV}(\vct{z}_{\tau+1}),
\end{align}
where $prox_f$ denotes the proximal mapping associated with function $f$. More details on Wirtinger-derivatives, its properties and applications to phase retrieval can be found in \cite{wf:candes2015phase}. The generalized gradient of $\mathcal{L}( \vct{x})$ takes the form
\begin{equation} \label{eq:grad}
\nabla \mathcal{L}(\vct{x}) = -\gcon \sum_{l=1}^{L} \Tl^H diag(\cgl) \mtx{A}^H(\mtx{A} \gl - \vct{y}_l \odot sgn(\mtx{A} \gl)),
\end{equation}
where $sgn(\cdot)$ denotes the complex signum function and $\odot$ stands for elementwise multiplication. 
We choose the step size
$
\mu_{\tau} = \frac{1}{\Gamma_{\tau}},
$
where 
\begin{align}\label{eq:Lt}
\Gamma_{\tau} = \frac{4 \pi^2}{\lambda^2} \Bigg[\sum_{\ell=1}^{L}  \left\|\sum_{k=1}^{K} diag(|\vct{p}_k|^2)~diag(|\gl|^2)\right\|_2 +\left\|diag\Big[\Big( \pdiff{\gl}\mathcal{L}(\vct{x}) \Big)^H \odot \cgl \Big]\right\|_2\Bigg].
\end{align}
We note that $\Gamma_{\tau}$ can be determined from the known probe and quantities computed whilst calculating the gradient (current exit wave estimate, gradient w.r.t. exit wave) and hence requires no additional effort. This step size is motivated by a theoretical bound on the spectral norm of the loss Hessian that describes the maximum variation of the loss landscape and works well in practice. In the next section, we provide formal convergence guarantees for a slightly more conservative step size.

Due to the ambiguities discussed in Section \ref{sec:model}, the loss landscape $\mathcal{L}(\vct{x})$ has many undesired global optima. Furthermore, due to the highly nonlinear nature of the forward model the loss is highly nonconvex with many local optima.
3D-AWF biases the optimization process towards the desired reconstruction by exploiting \textit{a priori} knowledge on the structure of the solution via TV proximal mappings. The benefit of TV-regularization is threefold: (1)  it expedites data acquisition time drastically through decreasing the necessary number of measurements required for accurate reconstruction, (2) it helps resolve the ambiguity introduced by phase wrapping to a high degree and (3) serves as excellent prior for integrated circuits due to their highly structured, piecewise constant nature.

In Section \ref{sec:model} we showed that there is a voxel-level ambiguity in the real part of the object due to phase wrapping. Applying 3D TV regularization promotes a piecewise constant structure over the 3D reconstruction. Since we know a priori that the ground truth object is piecewise constant, this in turn ensures that the ambiguity in $\vct{d}$ is also piecewise constant. Therefore, it opens up a way to mitigate the phase wrapping effect by facilitating the approximation of the ambiguity by a single constant over the object: $\widehat{\vct{d}} \approx \vct{d}^* + \bold{1} \widetilde{d}$ with $\widetilde{d} \in \mathbb{R}$. Finding the optimal constant $\widetilde{d}$ necessitates some knowledge on the ground truth object. We assume that some pixels of the ground truth exit waves are known, which translates to knowing some line integrals through the ground truth object. This information is readily available by using the part of the 3D object which is known to be vacuum or a given substrate. Denote $\mtx{D}_\ell$ the diagonal operator that masks out unknown pixels in the ground truth projection image in orientation $\ell$, so that $\mtx{D}_\ell \Tl d^*$ is known. Then we can obtain  $\widetilde{d}$ by solving 
\begin{align}\label{eq:mincorr}
\min_{\tilde{d}}~ \sum_{\ell=1}^L ||\mtx{D}_\ell \Tl\left(\widehat{\vct{d}} - \bold{1} \widetilde{d}\right) - \mtx{D}_\ell \Tl \vct{d}^*||_2^2,
\end{align}
for which the solution can be easily calculated in closed form by
\begin{equation*}\label{eq:corr1}
\tilde{d} = \frac{\bold{1}^H \sum_{\ell=1}^L \mtx{T}_\ell^H(\mtx{D}_\ell \mtx{T}_\ell \hat{\vct{d}}- \mtx{D}_\ell \mtx{T}_\ell \vct{d}^*)}{\bold{1}^H \sum_{\ell=1}^L \mtx{T}_\ell^H \mtx{D}_\ell \mtx{T}_\ell \bold{1}}.
\end{equation*}
Let $\widehat{\vct{x}}_T$ be the full reconstruction obtained from running 3D-AWF for $T$ iterations. Then, our correction technique yields the final reconstruction $ \widehat{\vct{x}}_F$ given by
 \begin{align}\label{eq:corr}
 \widehat{\vct{x}}_F = \widehat{\vct{x}}_T - \bold{1} \tilde{d}.
 \end{align}
	\section{Convergence theory}\label{apx:stepsize:}
The loss function in \eqref{eq:optprob} is non-differentiable and highly non-convex. Therefore it is completely unclear why 3D-AWF even converges. In the next theorem we ensure convergence to a stationary point. We defer the proof to Appendix \ref{apx:proof}.
\begin{theorem}\label{thm:main}
	Let $\vct{x} \in \mathbb{C}^N$ represent the object and assume we have noisy measurements of the form 
	$ \vct{y}_\ell = |\mtx{A} \gl | + \vct{n}_\ell$ corresponding to projection angles $\ell = 1,\ldots,L$. Here, $\gl \in \mathbb{C}^P$ is defined per \eqref{eq:gl} and $\vct{n}_\ell$ is used to denote arbitrary noise on the measurements from the $\ell$th angle. We run 3D-AWF updates of the form \eqref{eq:awfprox} with $\beta_\tau = 0$ with step size
	$$\mu \leq\left[ \frac{4\pi^2}{\lambda^2}\left( (1+ \sqrt{P}) L \lambda_{max} + \sqrt{ \lambda_{max}} \sum_{\ell =1}^{L} \twonorm{\vct{y}_\ell}\right) \right] ^{-1},$$
	where $ \lambda_{max} = \left\|\sum_{k=1}^{K} diag(|\vct{p}_k|^2)\right\|_2$. Furthermore, let $\vct{x}^*$ be a global optimum of $\mathcal{L}_{total}(\vct{x})$. Then, we have 
	\begin{equation*}
	\lim_{\tau \rightarrow \infty} \twonorm{prox_{TV}(\vct{z}_{\tau})- \vct{x}_{\tau}} = 0, 
	\end{equation*}
	and more specifically
	\begin{equation*}
	\min_{\tau \in \{1,2,...,T\}} \twonorm{prox_{TV} (\vct{z}_\tau) - \vct{x}_\tau}^2  \leq \mu \frac{\mathcal{L}_{total}(\vct{x}_0) - \mathcal{L}_{total}(\vct{x}^*)}{T+1}.
	\end{equation*}
\end{theorem}
This theorem guarantees that if we choose the step size smaller than a constant which can be calculated purely based on our measurements and the known probe function, then 3D-AWF will converge to a stationary point. Moreover, the norm of the difference of iterates diminishes proportional to $\frac{1}{T}$. It is important to note that even though Theorem \ref{thm:main}  is formulated in terms of TV regularization for this particular application, a more general result in Appendix \ref{apx:proof} shows that 3D-AWF converges for any convex regularizer.
	\section{Numerical experiments}
In this section, we investigate the performance of 3D-AWF in the context of ptychographic phaseless
imaging of 3D samples. We perform the reconstruction on a complex 3D test image of size $124 \times 124 \times 220$ voxels ($N \approx 3.4 \cdot 10^6$) obtained from a highly realistic synthetic IC structure specified in \cite{awf:xu2018accelerated}. We use a simulated x-ray source with an energy of $6.2 keV$ ($\lambda_0 = 0.2 nm$).
To generate the measurements we repeat the ptychographic acquisition procedure with parameters described in \cite{awf:xu2018accelerated} for $L = \{5, 10, 25, 50, 100, 250, 400\}$ different illumination angles, where the object is rotated by $\frac{\pi}{L}$ increments about its $y$ axis. 
\begin{algorithm}
	\caption{3D-AWF}\label{alg:awf}
	\begin{algorithmic}[1]
		\Require $\lambda_{TV}, \vct{y} := \{\vct{y}_\ell\}_{\ell=1,2,..,L}$
		\State $\hat{\vct{x}}_0 \gets 0$ \Comment{Initialization}
		\For{$\tau= 1 ~\text{to}~ T$}
		\State{$\beta_\tau \gets \frac{\tau+1}{\tau+3}$}
		\State{$\vct{q}_{\tau+1}  \gets \vct{x}_{\tau} + \beta_{\tau}(\vct{x}_{\tau} - \vct{x}_{\tau-1}) $} \Comment{Temporary variable}
		\State $\nabla \mathcal{L}( \vct{q}_{\tau+1}) \gets \text{gradient}(\vct{q}_{\tau+1}, \vct{y})$ \Comment{Gradient from \eqref{eq:grad}}
		\State $\mu_{\tau} \gets 1/ \Gamma_\tau$ \Comment{Step size from \eqref{eq:Lt}}
		\State{$\vct{z}_{\tau+1} \gets \vct{q}_{\tau+1} - \mu_{\tau} \nabla \mathcal{L}( \vct{q}_{\tau+1})$}
		\State{$\vct{x}_{\tau+1} \gets prox_{TV}(\vct{z}_{\tau+1})$}
		\EndFor
		\State $\hat{\vct{x}}_F = \text{correction}(\vct{x}_T)$ \Comment{Correction from \eqref{eq:corr}}
		\Ensure $\hat{\vct{x}}_F$ \Comment{Final reconstruction}
	\end{algorithmic}
\end{algorithm}
First, we implement 3D-AWF (Algorithm \ref{alg:awf}) to minimize the TV-regularized problem defined in \eqref{eq:optprob} with iterative proximal update rule in \eqref{eq:awfprox} with $T=550$ iterations. 
We tune the regularization strength by minimizing reconstruction error with $L=100$ illumination angles. For experiments with different number of angles we scale the regularizer linearly with $L$ to maintain the ratio of TV-penalty to the total loss. The tuned value for 3D-AWF at $100$ angles is $\vct{\lambda}^{AWF}_{TV} = 0.1$.
The chip has a fine, layered structure along the $z$-axis, therefore we set the regularization weights to $\vct{w} = [1, 1, 0.1]$ to enforce a piecewise constant structure mostly in the $x-y$ plane. We report the relative error on the corrected reconstruction (output of Algorithm \ref{alg:awf}) as
$
RE_{final} =\left\|\mathbf{M}\left(\widehat{\vct{x}}_F-\vct{x}^*\right)\right\|_2 / \left\|\mathbf{M}\mathbf{x}^*\right\|_2.
$
Here, $\mtx{M}$ extracts the center $62 \times 62 \times 110$ voxel region of the object (the region-of-interest), outside of which the object did not receive enough illumination from the probes and therefore we don't expect to have accurate reconstruction in that region. 

We compare our results to a combined, two-step (2-Step) approach in which we first perform 2D phase retrieval then reconstruct the object from projections via tomography. In the first step, we reconstruct the exit waves by minimizing
$
\sum_{\ell=1}^{L} \|\vct{y}_{\ell}- |\mtx{A} \vct{f}_\ell|\|_2^2,
$
yielding estimated exit waves $\widehat{\vct{f}}_\ell$. 
In this method the exit wave is approximated based on its Taylor-series expansion as 
$
exp(\gcon \Tl \vct{x}) \approx 1 + \gcon \Tl \vct{x}
$
yielding the loss function for tomography
\begin{align}\label{eq:lossfbp2}
\sum_{\ell=1}^{L} \|  \mtx{H}\widehat{\vct{f}}_\ell -\gcon\mtx{H}\Tl \vct{x}\|_2^2+	\lambda_{TV} \textbf{TV}_{3D}(\vct{x}),
\end{align}
where $\mtx{H}$ represents the ramp filter used in filtered backprojection aimed at inverting the Radon transform. 
We will assume that the exit waves have been reconstructed perfectly in the phase retrieval step (that is $\widehat{\vct{f}}_\ell = \vct{f}^*_\ell,~ \ell = 1, 2,..,L$) and run conjugate gradient descent on the loss function in \eqref{eq:lossfbp2} for $T=550$ iterations. We tune and scale the regularizer by the same methodology as for 3D-AWF with $\vct{\lambda}^{2-Step}_{TV} = 10^4$ for $L=100$ angles. To perform the correction, we assume that the same pixel values are known as in case of 3D-AWF and report the final reconstruction error after correction.

\begin{figure}
	\centering
	\begin{minipage}{.49\textwidth}
		\centering
		\includegraphics[width=0.85\linewidth]{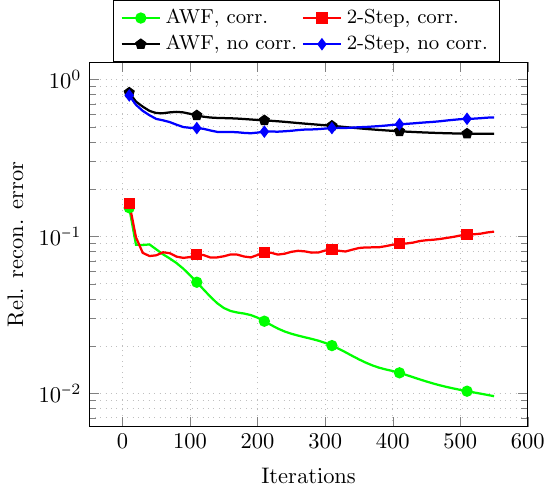}
		\captionof{figure}{Evolution of relative reconstruction error before and after correction across iterations. $L=100.$}
		\label{fig:rel_err_t}
	\end{minipage}
	\begin{minipage}{.49\textwidth}
		\centering
		\includegraphics[width=0.83\linewidth]{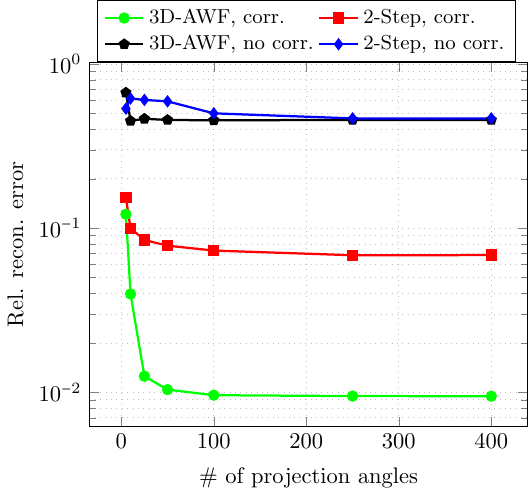}
		\captionof{figure}{Relative reconstruction error before and after correction vs. number of angles. }
		\label{fig:rel_err}
	\end{minipage}
\end{figure}

In case of 2-Step, we observe that the relative reconstruction error achieves its minimum fairly early (100-150 iterations) and increases afterwards, with consistently worse reconstructions at iteration $550$. Therefore we show the best reconstruction across all iterations for this technique. On the other hand, as it
is observed on Fig. \ref{fig:rel_err_t}, 3D-AWF reconstruction error is decreasing throughout iterations and therefore we report results for the last iteration. 3D-AWF reconstruction improves with more iterations, which cannot be said for 2-Step.
We note that the relative error before correction is consistently high for both algorithms. We attribute this fact to the inherent ambiguity of the reconstruction problem, which emphasizes the need to incorporate some form of prior knowledge. After applying the correction technique described in \eqref{eq:corr}, the reconstruction error decreases drastically for both algorithms.

\begin{figure}
	\centering
	\includegraphics[scale=1.30]{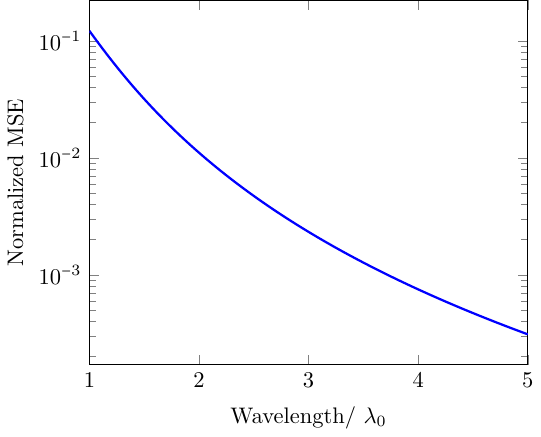}
	\caption{Difference between the exponential model and its linear approximation. We plot the normalized mean squared error between an exit wave obtained from the non-linear model and the linearized model at various wavelengths (normalized by the wavelength used in the experiment). At high energies (short wavelengths) the linear approximation significantly deviates from the exponential model. \label{fig:lambdas}}
\end{figure}

\begin{figure}
	\centering
	\includegraphics[scale=0.30]{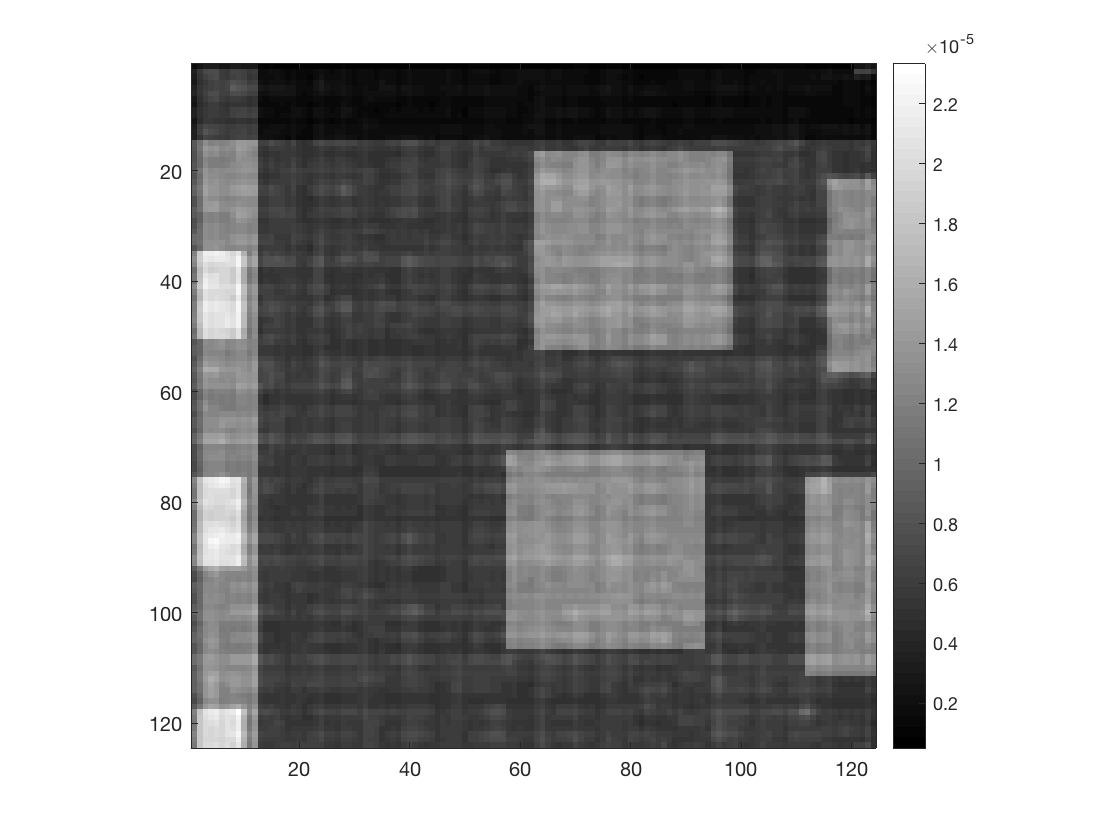}
	\caption{Normalized pixelwise squared difference between exit waves calculated from the exponential propagation model and the linearized model. The error is significantly higher at pixels resulting from the illuminating beam passing through metallic parts, such as copper interconnects in the object.\label{fig:lin_error_img}}
\end{figure}

Fig. \ref{fig:rel_err} depicts relative reconstruction error achieved by 3D-AWF and 2-Step for various number of illumination angles. These results show that 3D-AWF achieves significantly better reconstruction accuracy with significantly fewer angles. We attribute most of this difference to the inaccuracy of the linear model used in 2-Step. Fig. \ref{fig:lambdas} shows how the linear approximation increasingly deviates from the exponential model at shorter wavelengths, such as the one used in our simulation. Imaging with high energy beams (or short wavelengths) is crucial for obtaining nano-scale resolution. Moreover, our experiments show that the presence of metallic parts in the object further increases the inaccuracy of the linear model (Fig. \ref{fig:lin_error_img}). This is due to the fact that metals typically have high attenuation (represented by $\vct{b}$ in Section \ref{sec:model}, the imaginary part of the complex refractive index). All these observations highlight the advantage of the exponential model over the linear approximation for high resolution imaging of integrated circuits of significant spatial extent.

\begin{figure}
	\centering
	\includegraphics[scale=0.215]{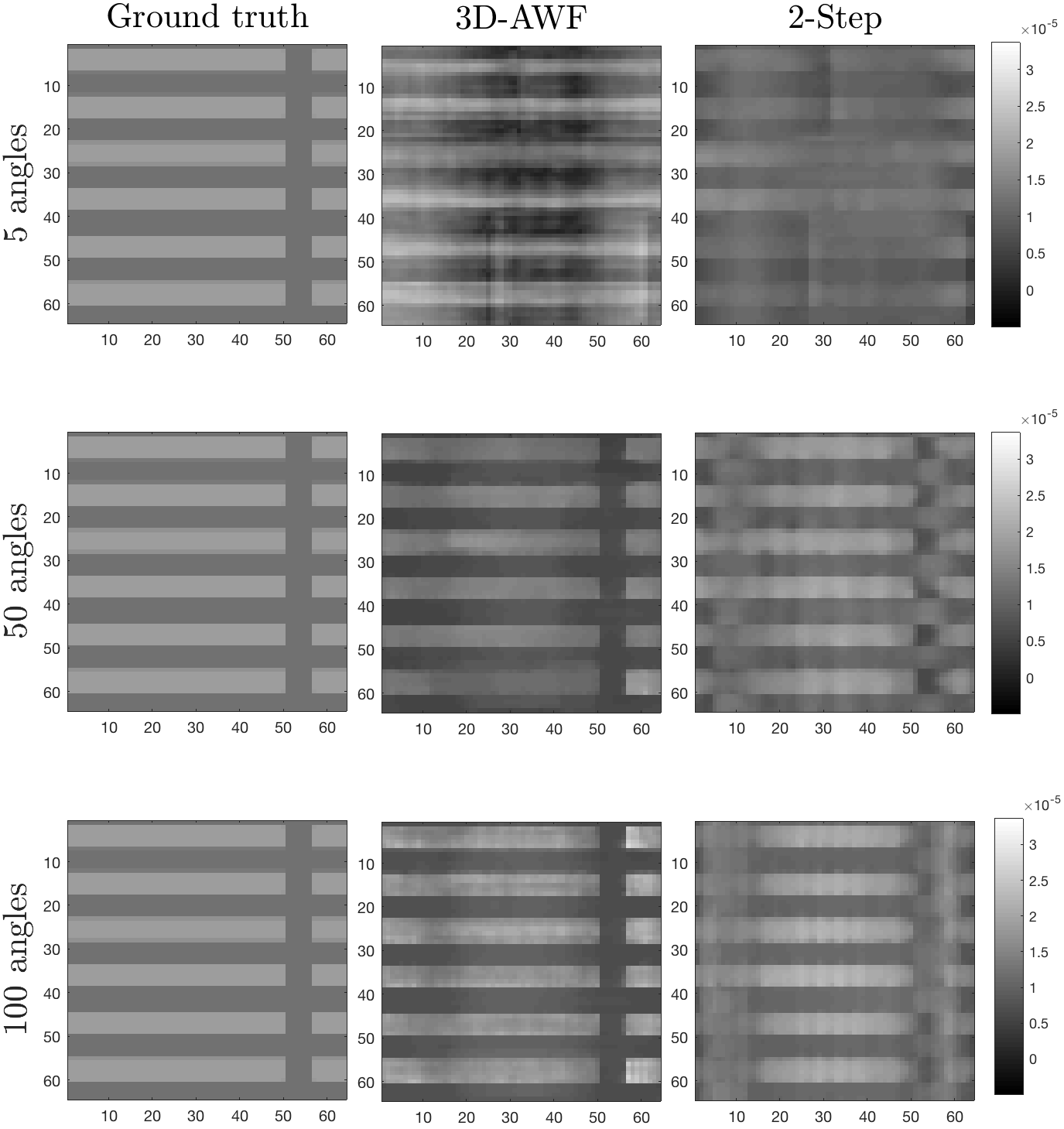}
	\caption{Magnitude of ground truth of a slice ($x-y$ plane at $z=1$) of 3D-AWF and 2-Step reconstructions after correction.  \label{fig:recon}}
\end{figure}
\begin{figure}
	\centering
	\includegraphics[scale=0.3]{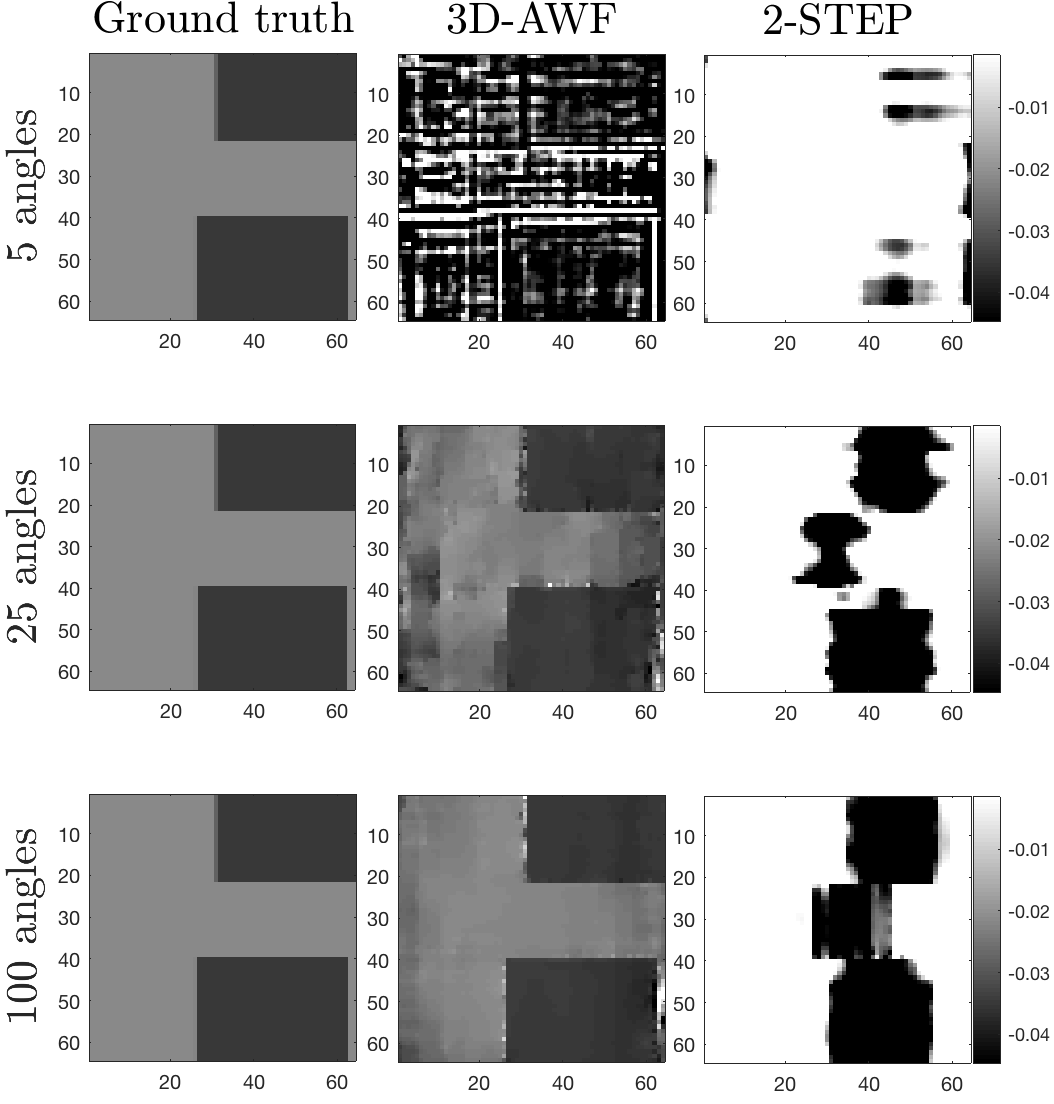}
	\caption{Phase of ground truth of a slice ($x-y$ plane at $z=60$) of 3D-AWF and 2-Step reconstructions after correction.  \label{fig:recon_phase}}
\end{figure}
\begin{figure}
	\centering
	\includegraphics[scale=0.235]{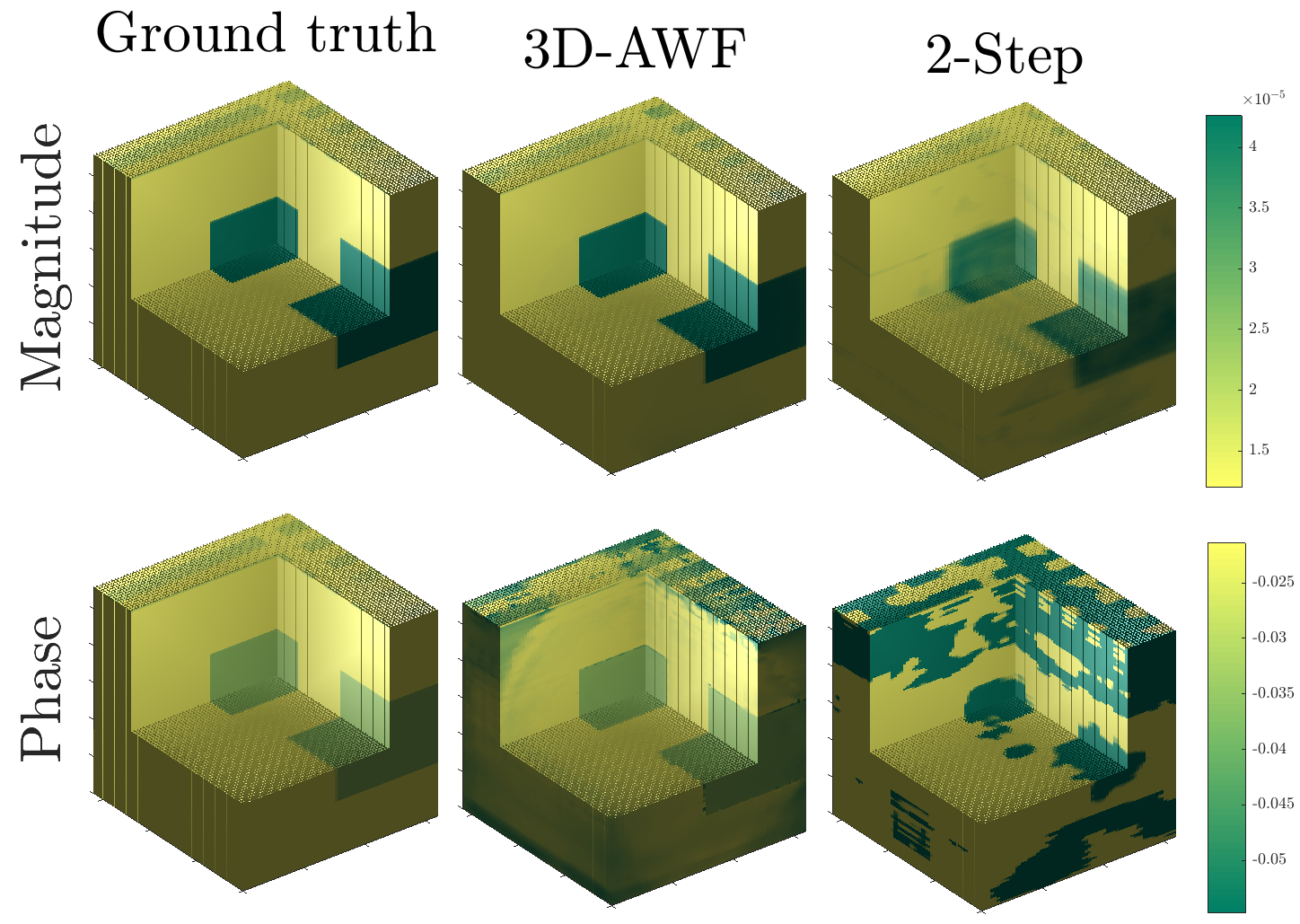}
	\caption{3D rendering of the magnitude and phase of the ground truth and reconstructed volumes using 3D-AWF and 2-Step ($L = 100$).  \label{fig:3drecon}}
\end{figure}

Lastly, we plot the magnitude of a slice of the ground truth object and reconstructions after correction in Fig. \ref{fig:recon} for various projection angles. Even though the reconstructions significantly improve with more illumination angles, visible reconstruction quality saturates after $100$ angles.  Reconstruction of the magnitude image using 3D-AWF is highly accurate with sharp edges even with low number of measurements. Edges on the 2-Step magnitude plot are less well-defined and magnitude values are inaccurate.  The phase plots (Figure \ref{fig:recon_phase}) show drastic differences between the two reconstruction algorithms. In general, the phase of the object converges significantly slower than the magnitude and is less accurate, which is due to the loss of phase information in the measurement process. One may observe that the phase plot of 2-Step exhibit serious inaccuracies, even after correction. A 3D rendering of the reconstructed volume using $L=100$ illumination angles can be seen on Figure \ref{fig:3drecon}. The quality of 3D-AWF reconstruction is visibly better throughout the volume, and we observe lower reconstruction error close to the center of the object due to the geometry of the setup.

	\clearpage
	\bibliographystyle{IEEEbib}
	\small{
	\bibliography{strings,refs}

\begin{thebibliography}{10}

\bibitem{phaseless:abbey2008keyhole}
Brian Abbey, Keith~A Nugent, Garth~J Williams, Jesse~N Clark, Andrew~G Peele,
  Mark~A Pfeifer, Martin De~Jonge, and Ian McNulty,
\newblock ``Keyhole coherent diffractive imaging,''
\newblock {\em Nature Physics}, 2008.

\bibitem{phaseless:clark2013ultrafast}
JN~Clark, L~Beitra, G~Xiong, A~Higginbotham, DM~Fritz, HT~Lemke, D~Zhu,
  M~Chollet, GJ~Williams, Marc Messerschmidt, et~al.,
\newblock ``Ultrafast three-dimensional imaging of lattice dynamics in
  individual gold nanocrystals,''
\newblock {\em Science}, 2013.

\bibitem{phaseless:miao1999extending}
Jianwei Miao, Pambos Charalambous, Janos Kirz, and David Sayre,
\newblock ``Extending the methodology of x-ray crystallography to allow imaging
  of micrometre-sized non-crystalline specimens,''
\newblock {\em Nature}, 1999.

\bibitem{phaseless:nelson2010high}
Johanna Nelson, Xiaojing Huang, Jan Steinbrener, David Shapiro, Janos Kirz,
  Stefano Marchesini, Aaron~M Neiman, Joshua~J Turner, and Chris Jacobsen,
\newblock ``High-resolution x-ray diffraction microscopy of specifically
  labeled yeast cells,''
\newblock {\em Proceedings of the National Academy of Sciences}, 2010.

\bibitem{phaseless:pfeifer2006three}
Mark~A Pfeifer, Garth~J Williams, Ivan~A Vartanyants, Ross Harder, and Ian~K
  Robinson,
\newblock ``Three-dimensional mapping of a deformation field inside a
  nanocrystal,''
\newblock {\em Nature}, 2006.

\bibitem{phaseless:shapiro2005biological}
David Shapiro, Pierre Thibault, Tobias Beetz, Veit Elser, Malcolm Howells,
  Chris Jacobsen, Janos Kirz, Enju Lima, Huijie Miao, Aaron~M Neiman, et~al.,
\newblock ``Biological imaging by soft x-ray diffraction microscopy,''
\newblock {\em Proceedings of the National Academy of Sciences}, 2005.

\bibitem{ptychoexp:deng2015simultaneous}
Junjing Deng, David~J Vine, Si~Chen, Youssef~SG Nashed, Qiaoling Jin,
  Nicholas~W Phillips, Tom Peterka, Rob Ross, Stefan Vogt, and Chris~J
  Jacobsen,
\newblock ``Simultaneous cryo x-ray ptychographic and fluorescence microscopy
  of green algae,''
\newblock {\em Proceedings of the National Academy of Sciences}, 2015.

\bibitem{ptychoexp:dierolf_nature}
Martin Dierolf, Andreas Menzel, Pierre Thibault, Philipp Schneider, Cameron~M.
  Kewish, Roger Wepf, Oliver Bunk, and Franz Pfeiffer,
\newblock ``Ptychographic x-ray computed tomography at the nanoscale,''
\newblock {\em Nature}, 2010.

\bibitem{ptychoexp:holler2017high}
Mirko Holler, Manuel Guizar-Sicairos, Esther~HR Tsai, Roberto Dinapoli,
  Elisabeth M{\"u}ller, Oliver Bunk, J{\"o}rg Raabe, and Gabriel Aeppli,
\newblock ``High-resolution non-destructive three-dimensional imaging of
  integrated circuits,''
\newblock {\em Nature}, 2017.

\bibitem{ptychoexp:shapiro2014chemical}
David~A Shapiro, Young-Sang Yu, Tolek Tyliszczak, Jordi Cabana, Rich Celestre,
  Weilun Chao, Konstantin Kaznatcheev, AL~David Kilcoyne, Filipe Maia, Stefano
  Marchesini, et~al.,
\newblock ``Chemical composition mapping with nanometre resolution by soft
  x-ray microscopy,''
\newblock {\em Nature Photonics}, 2014.

\bibitem{ptychoexp:thibault2008high}
Pierre Thibault, Martin Dierolf, Andreas Menzel, Oliver Bunk, Christian David,
  and Franz Pfeiffer,
\newblock ``High-resolution scanning x-ray diffraction microscopy,''
\newblock {\em Science}, 2008.

\bibitem{fourierptychoexp:horstmeyer2015digital}
Roarke Horstmeyer, Xiaoze Ou, Guoan Zheng, Phil Willems, and Changhuei Yang,
\newblock ``Digital pathology with fourier ptychography,''
\newblock {\em Computerized Medical Imaging and Graphics}, 2015.

\bibitem{fourierptychoexp:tian20153d}
Lei Tian and Laura Waller,
\newblock ``3d intensity and phase imaging from light field measurements in an
  led array microscope,''
\newblock {\em optica}, 2015.

\bibitem{fourierptychoexp:tian2015computational}
Lei Tian, Ziji Liu, Li-Hao Yeh, Michael Chen, Jingshan Zhong, and Laura Waller,
\newblock ``Computational illumination for high-speed in vitro fourier
  ptychographic microscopy,''
\newblock {\em Optica}, 2015.

\bibitem{fourierptychoexp:zheng2013wide}
Guoan Zheng, Roarke Horstmeyer, and Changhuei Yang,
\newblock ``Wide-field, high-resolution fourier ptychographic microscopy,''
\newblock {\em Nature photonics}, 2013.

\bibitem{partcohptychoexp:chang2018partially}
Huibin Chang, Pablo Enfedaque, Yifei Lou, and Stefano Marchesini,
\newblock ``Partially coherent ptychography by gradient decomposition of the
  probe,''
\newblock {\em Acta Crystallographica Section A: Foundations and Advances},
  2018.

\bibitem{awf:xu2018accelerated}
Rui Xu, Mahdi Soltanolkotabi, Justin~P Haldar, Walter Unglaub, Joshua Zusman,
  Anthony~FJ Levi, and Richard~M Leahy,
\newblock ``Accelerated wirtinger flow: A fast algorithm for ptychography,''
\newblock {\em arXiv preprint arXiv:1806.05546}, 2018.

\bibitem{wf:candes2015phase}
Emmanuel~J Candes, Xiaodong Li, and Mahdi Soltanolkotabi,
\newblock ``Phase retrieval via wirtinger flow: Theory and algorithms,''
\newblock {\em IEEE Transactions on Information Theory}, 2015.

\bibitem{pradditional:chandra2017phasepack}
Rohan Chandra, Ziyuan Zhong, Justin Hontz, Val McCulloch, Christoph Studer, and
  Tom Goldstein,
\newblock ``Phasepack: A phase retrieval library,''
\newblock in {\em 2017 51st Asilomar Conference on Signals, Systems, and
  Computers}. IEEE, 2017.

\bibitem{pradditional:waldspurger2018phase}
Ir{\`e}ne Waldspurger,
\newblock ``Phase retrieval with random gaussian sensing vectors by alternating
  projections,''
\newblock {\em IEEE Transactions on Information Theory}, 2018.

\bibitem{pradditional:jaganathan2012recovery}
Kishore Jaganathan, Samet Oymak, and Babak Hassibi,
\newblock ``Recovery of sparse 1-d signals from the magnitudes of their fourier
  transform,''
\newblock in {\em 2012 IEEE International Symposium on Information Theory
  Proceedings}. IEEE, 2012.

\bibitem{pradditional:soltanolkotabi2019structured}
Mahdi Soltanolkotabi,
\newblock ``Structured signal recovery from quadratic measurements: Breaking
  sample complexity barriers via nonconvex optimization,''
\newblock {\em IEEE Transactions on Information Theory}, 2019.

\bibitem{ptycho3d:maiden2012multisclice}
Andrew~M Maiden, Martin~J Humphry, and JM~Rodenburg,
\newblock ``Ptychographic transmission microscopy in three dimensions using a
  multi-slice approach,''
\newblock {\em JOSA A}, 2012.

\bibitem{ptycho3d:tian20153d}
Lei Tian and Laura Waller,
\newblock ``3d intensity and phase imaging from light field measurements in an
  led array microscope,''
\newblock {\em optica}, 2015.

\bibitem{lamino3d:myagotin2013efficient}
Anton Myagotin, Alexey Voropaev, Lukas Helfen, Daniel H{\"a}nschke, and Tilo
  Baumbach,
\newblock ``Efficient volume reconstruction for parallel-beam computed
  laminography by filtered backprojection on multi-core clusters,''
\newblock {\em IEEE Transactions on Image Processing}, 2013.

\bibitem{pct:gursoy2017direct}
Do{\u{g}}a G{\"u}rsoy,
\newblock ``Direct coupling of tomography and ptychography,''
\newblock {\em Optics letters}, 2017.

\bibitem{pct:aslan2019joint}
Selin Aslan, Viktor Nikitin, Daniel~J Ching, Tekin Bicer, Sven Leyffer, and
  Do{\u{g}}a G{\"u}rsoy,
\newblock ``Joint ptycho-tomography reconstruction through alternating
  direction method of multipliers,''
\newblock {\em Optics express}, 2019.

\bibitem{pct:nikitin2019photon}
Viktor Nikitin, Selin Aslan, Yudong Yao, Tekin Bi{\c{c}}er, Sven Leyffer,
  Rajmund Mokso, and Do{\u{g}}a G{\"u}rsoy,
\newblock ``Photon-limited ptychography of 3d objects via bayesian
  reconstruction,''
\newblock {\em OSA Continuum}, 2019.

\bibitem{pct:chang2019iterative}
Huibin Chang, Pablo Enfedaque, and Stefano Marchesini,
\newblock ``Iterative joint ptychography-tomography with total variation
  regularization,''
\newblock {\em arXiv preprint arXiv:1902.05647}, 2019.

\bibitem{book:clarke1990optimization}
Frank~H Clarke,
\newblock {\em Optimization and nonsmooth analysis}, vol.~5,
\newblock Siam, 1990.

\end{thebibliography}
	}

	\appendix
\section{Proof of  Theorem \ref{thm:main}}\label{apx:proof}
Here, we are going to prove our main result on the convergence of 3D-AWF to stationary points stated in Theorem \ref{thm:main}. We are going to use Wirtinger-derivatives in place of regular differentiation. For an overview on the notion of Wirtinger-derivatives and some properties we refer the reader to \cite{wf:candes2015phase}. Let $\bar{x}$ denote the complex conjugate of $x \in \mathbb{C}$ and for a matrix $\mtx{A} \in \mathbb{C}^{n \times m}$ we write $\mtx{A}^H =\bar{ \mtx{A}}^T \in \mathbb{C}^{m \times n}$ its Hermitian transpose. 

First, we want to upperbound the spectral norm of the Hessian of $\mathcal{L}(x)$. Let
\begin{equation*}
\jac\gl = \pdiff{\vct{x}} \gl
\end{equation*}
denote the Jacobian of $\gl$.
Since
\begin{equation*}
\{ \jac\gl \}_{i,j} = \gcon \{\Tl\}_{i,j} \{\gl\}_i,
\end{equation*}
and therefore we have 
\begin{equation}\label{eq:jac}
\jac\gl = \gcon \Tl \odot  [ \gl~ \gl~ \gl~ .. ~\gl] = \gcon diag(\gl) \Tl.
\end{equation}
Note that the "mixed" derivatives 
\begin{equation*}
\pdiff{\cx} \gl = 0,~~ \pdiff{x} \cgl = \pdiff{\vct{x}} e^{-\gcon \Tl \cx} = 0.
\end{equation*}
Moreover 
\begin{equation*}
\overline{\jac \gl}(x) = \pdiff{\cx} \cgl = -\gcon diag(\cgl) \Tl^H.
\end{equation*}
Therefore, the complex gradient of the loss function takes the form
\begin{align} \label{grad}
\nabla \mathcal{L}(\vct{x}) &= \sum_{l=1}^{L} \jac\gl^H \mtx{A}^H(\mtx{A} \gl - \vct{y}_\ell \odot sgn(\mtx{A} \gl)) \nonumber\\ 
&= -\gcon \sum_{l=1}^{L} \Tl^H diag(\cgl) \mtx{A}^H(\mtx{A} \gl - \vct{y}_\ell \odot sgn(\mtx{A} \gl)).
\end{align}
To find the Hessian, first consider the smoothed 1D problem in the form
\begin{equation}\label{loss1d}
\mathcal{L}_{\epsilon}(\vct{x}) = \sum_{l=1}^{L} \sum_{m=1}^{M/L}\Big(\left(|\vct{a}_m^H \gl|^2 + \epsilon\right)^{\frac{1}{2}}-y_{m,l}\Big)^2,
\end{equation}
where $\vct{a}_m$ represents the $m^{th}$ row of $\mtx{A}$ as a column vector and $y_{m,l}$ is the $m^{th}$ entry of $\vct{y}_\ell$.
Rewriting \eqref{loss1d} as a holomorphic function of $\gl$ and its conjugate, we obtain
\begin{equation*}
\Big( \pdiff{\vct{x}}\mathcal{L}_{\epsilon}(\vct{x}) \Big)^T = \sum_{l=1}^{L} \sum_{m=1}^{M/L}\frac{\left(\gl^T(\vct{a}_m \vct{a}_m^H)^T \cgl + \epsilon\right)^{\frac{1}{2}}-y_{m,l}}{\left(\gl^T(\vct{a}_m \vct{a}_m^H)^T \cgl + \epsilon\right)^{\frac{1}{2}}} \jac\gl^T(\vct{a}_m \vct{a}_m^H)^T\cgl,
\end{equation*}
and therefore by substituting the Jacobian from \eqref{eq:jac} we have
\begin{align*}
\Big( \pdiff{\vct{x}}\mathcal{L}_{\epsilon}(\vct{x}) \Big)^H = -\gcon \sum_{l=1}^{L} \sum_{m=1}^{M/L}\frac{\left(|\vct{a}_m^H \gl|^2 + \epsilon\right)^{\frac{1}{2}}-y_{m,l}}{\left(|\vct{a}_m^H \gl|^2 + \epsilon\right)^{\frac{1}{2}}}
\Tl^H diag(\cgl)(\vct{a}_m \vct{a}_m^H)\gl,
\end{align*}
Now, applying the chain rule we obtain the second derivatives as
\begin{align*}
\mtx{H}_{gg} &= \pdiff{\vct{x}}\Big( \pdiff{\vct{x}}\mathcal{L}_{\epsilon}(\vct{x}) \Big)^H \\
&= \sum_{l=1}^{L} \sum_{m=1}^{M/L} \Big[ \frac{\left(|\vct{a}_m^H \gl|^2 + \epsilon\right)^{\frac{1}{2}}-y_{m,l}}{\left(|\vct{a}_m^H \gl|^2 + \epsilon\right)^{\frac{1}{2}}} +\frac{1}{2}\frac{|\vct{a}_m^H \gl|^2 ~y_{m,l}}{\left(|\vct{a}_m^H \gl|^2 + \epsilon\right)^{\frac{3}{2}}} \Big] \jac\gl^H(\vct{a}_m \vct{a}_m^H)\jac\gl\\
&=\frac{4\pi}{\lambda^2}\sum_{l=1}^{L} \sum_{m=1}^{M/L} \Big[ 1- \frac{y_{m,l}}{\left(|\vct{a}_m^H \gl|^2 + \epsilon\right)^{\frac{1}{2}}} +\frac{1}{2}\frac{|\vct{a}_m^H \gl|^2 ~y_{m,l}}{\left(|\vct{a}_m^H \gl|^2 + \epsilon\right)^{\frac{3}{2}}} \Big]\\
&\cdot \Tl^H diag(\cgl)(\vct{a}_m \vct{a}_m^H)diag(\gl)\Tl
\end{align*}

\begin{align*}
\mtx{H}_{\bar{g}g} &= \pdiff{\cx}\Big( \pdiff{\vct{x}}\mathcal{L}_{\epsilon}(\vct{x}) \Big)^H \\
&= \pdiff{\cx}\Big(-\gcon \sum_{l=1}^{L} \sum_{m=1}^{M/L}\frac{\left(|\vct{a}_m^H \gl|^2 + \epsilon\right)^{\frac{1}{2}}-y_{m,l}}{\left(|\vct{a}_m^H \gl|^2 + \epsilon\right)^{\frac{1}{2}}}
\Tl^H diag[(\vct{a}_m \vct{a}_m^H)\gl] \cgl\Big)\\
&= \sum_{l=1}^{L} \sum_{m=1}^{M/L} \frac{1}{2} \frac{(\vct{a}_m^H \gl)^2  y_{m,l}}{\left(|\vct{a}_m^H \gl|^2 + \epsilon\right)^{\frac{3}{2}}} \jac\gl^H(\vct{a}_m \vct{a}_m^T)\overline{\jac\gl} \\
&-\gcon \sum_{l=1}^{L} \sum_{m=1}^{M/L} \Big( 1- \frac{y_{m,l}}{\left(|\vct{a}_m^H \gl|^2 + \epsilon\right)^{\frac{1}{2}}}\Big)\Tl^H diag[(\vct{a}_m \vct{a}_m^H)\gl]\overline{\jac\gl}
\end{align*}

To find the largest singular value of the Hessian we want to upper bound the quadratic form
\begin{equation*}
\begin{bmatrix}
\vct{u} \\
\bar{\vct{u}}
\end{bmatrix}^H \nabla^2\mathcal{L}_{\epsilon}(\vct{x})
\begin{bmatrix}
\vct{u} \\
\bar{\vct{u}} 
\end{bmatrix} = \vct{u}^H \mtx{H}_{gg}\vct{u}+u^H \mtx{H}_{\bar{g}g} \bar{u}+\vct{u}^T \mtx{H}_{g\bar{g}}\vct{u} + \vct{u}^T \mtx{H}_{\bar{g}\bar{g}}\bar{\vct{u}}. 
\end{equation*}

The first term takes the form
\begin{align*}
\vct{u}^H \mtx{H}_{gg}\vct{u} &= \sum_{l=1}^{L} \sum_{m=1}^{M/L} \Big( 1- \frac{1}{2}\frac{y_{m,l}}{\left(|\vct{a}_m^H \gl|^2 + \epsilon\right)^{\frac{1}{2}}} - \frac{\epsilon}{2}\frac{y_{m,l}}{\left(|\vct{a}_m^H \gl|^2 + \epsilon\right)^{\frac{3}{2}}}\Big)|\vct{a}_m^H \jac\gl \vct{u}|^2.
\end{align*}

For the mixed terms we have
\begin{align*}
\vct{u}^H \mtx{H}_{\bar{g}g}\bar{\vct{\vct{u}}}+\vct{\vct{u}}^T \mtx{H}_{g\bar{g}}\vct{u} &= 2\Re \Bigg(\vct{u}^H \mtx{H}_{\bar{g}g}\bar{\vct{u}}\Bigg)\\
&=\sum_{l=1}^{L} \sum_{m=1}^{M/L} \frac{y_{m,l}}{\left(|\vct{a}_m^H \gl|^2 + \epsilon\right)^{\frac{3}{2}}} \Re\Bigg((\vct{a}_m^H \gl)^2 (\vct{u}^H \jac\gl^H\vct{a}_m)^2\Bigg)\\
&-\frac{8\pi^2}{\lambda^2}\sum_{l=1}^{L} \sum_{m=1}^{M/L} \Big( 1- \frac{y_{m,l}}{\left(|\vct{a}_m^H \gl|^2 + \epsilon\right)^{\frac{1}{2}}}\Big) \Re\Bigg((\vct{u}^H \Tl^H diag\Big[(\vct{a}_m \vct{a}_m^H)\gl \odot \cgl\Big] \Tl \bar{\vct{u}}\Bigg)\\
&=\sum_{l=1}^{L} \sum_{m=1}^{M/L} \frac{y_{m,l}}{\left(|\vct{a}_m^H \gl|^2 + \epsilon\right)^{\frac{3}{2}}} \Re\Bigg((\vct{a}_m^H \gl)^2 (\vct{u}^H \jac\gl^H\vct{a}_m)^2\Bigg)\\
&-\frac{8\pi^2} {\lambda^2}\sum_{l=1}^{L}\Re \Bigg(\vct{u}^H \Tl^Hdiag\Big[\Big( \pdiff{\gl}\mathcal{L}_{\epsilon}(\vct{x}) \Big)^H \odot \cgl \Big]\overline{\Tl}\bar{\vct{u}}\Bigg).
\end{align*}
Therefore,
\begin{align}
&\begin{bmatrix}
\vct{u} \\
\bar{\vct{u}}
\end{bmatrix}^H \nabla^2\mathcal{L}_{\epsilon}(\vct{x})
\begin{bmatrix}
\vct{u} \\
\bar{\vct{u}} 
\end{bmatrix} = \nonumber\\
&\sum_{l=1}^{L} \sum_{m=1}^{M/L} \Big( 1- \frac{1}{2}\frac{y_{m,l}}{\left(|\vct{a}_m^H \gl|^2 + \epsilon\right)^{\frac{1}{2}}} - \frac{\epsilon}{2}\frac{y_{m,l}}{\left(|\vct{a}_m^H \gl|^2 + \epsilon\right)^{\frac{3}{2}}}\Big)|\vct{a}_m^H \jac\gl \vct{u}|^2 \nonumber\\
&+\sum_{l=1}^{L} \sum_{m=1}^{M/L} \frac{y_{m,l}}{\left(|\vct{a}_m^H \gl|^2 + \epsilon\right)^{\frac{3}{2}}} \Re\Bigg((\vct{a}_m^H \gl)^2 (\vct{u}^H \jac\gl^H\vct{a}_m)^2\Bigg) \nonumber\\
& -\frac{8\pi^2} {\lambda^2}\sum_{l=1}^{L}\Re \Bigg(\vct{u}^H \Tl^Hdiag\Big[\Big( \pdiff{\gl}\mathcal{L}_{\epsilon}(\vct{x}) \Big)^H \odot \cgl \Big]\overline{\Tl}(x)\bar{\vct{u}}\Bigg) \nonumber\\
&=\sum_{l=1}^{L} \sum_{m=1}^{M/L} \Big( 1-\epsilon\frac{y_{m,l}}{\left(|\vct{a}_m^H \gl|^2 + \epsilon\right)^{\frac{3}{2}}}\Big)|\vct{a}_m^H \jac\gl \vct{u}|^2 \nonumber\\
&+\sum_{l=1}^{L} \sum_{m=1}^{M/L} \frac{y_{m,l}}{\left(|\vct{a}_m^H \gl|^2 + \epsilon\right)^{\frac{3}{2}}} \Re\Bigg((\vct{a}_m^H \gl)^2 (\vct{u}^H \jac\gl^H\vct{a}_m)^2 -|\vct{a}_m^H \gl|^2 |\vct{u}^H \jac\gl^H\vct{a}_m|^2 \Bigg) \nonumber\\
& -\frac{8\pi^2} {\lambda^2}\sum_{l=1}^{L}\Re \Bigg(\vct{u}^H \Tl^Hdiag\Big[\Big( \pdiff{\gl}\mathcal{L}_{\epsilon}(\vct{x}) \Big)^H \odot \cgl \Big]\overline{\Tl}\bar{\vct{u}}\Bigg) \nonumber\\
& \leq 2\sum_{l=1}^{L} \sum_{m=1}^{M/L} |\vct{a}_m^H \jac\gl \vct{u}|^2  -\frac{8\pi^2} {\lambda^2}\sum_{l=1}^{L}\Re \Bigg(\vct{u}^H \Tl^Hdiag\Big[\Big( \pdiff{\gl}\mathcal{L}_{\epsilon}(\vct{x}) \Big)^H \odot \cgl \Big]\overline{\Tl}\bar{\vct{u}}\Bigg) \nonumber\\
&= 2\sum_{l=1}^{L}\Bigg[\vct{u}^H \jac\gl^H \Big( \sum_{m=1}^{M/L}\vct{a}_m \vct{a}_m^H \Big)\jac\gl \vct{u} \nonumber\\
&-\frac{4\pi^2} {\lambda^2}\sum_{l=1}^{L}\Re \Bigg(\vct{u}^H \Tl^Hdiag\Big[\Big( \pdiff{\gl}\mathcal{L}_{\epsilon}(\vct{x}) \Big)^H \odot \cgl \Big]\overline{\Tl}\bar{\vct{u}}\Bigg) \Bigg] \label{qform}
\end{align}
Note that the diagonal matrix in the second term $\mtx{D}_\ell = diag\Big[\Big( \pdiff{\gl}\mathcal{L}_{\epsilon}(\vct{x}) \Big)^H \odot \cgl \Big]$ is directly calculated in each iteration, since it is the gradient corresponding to a certain angle before applying the adjoint operator $\Tl^H$.

Focusing on the first term in Eq. \eqref{qform} and letting $\mtx{P} = \sum_{k=1}^{K} diag(\vct{p}_k)^H diag(\vct{p}_k), $ a PSD diagonal matrix, we obtain
\begin{align*}
\sum_{l=1}^{L}\vct{u}^H \jac\gl^H \Big( \sum_{m=1}^{M/L}\vct{a}_m \vct{a}_m^H \Big)\jac\gl \vct{u} &=\frac{4 \pi^2}{\lambda^2}\sum_{l=1}^{L}\vct{u}^H \Tl^H diag(\gl) ^H \mtx{P}~ diag(\gl) \Tl \vct{u} \\
&\leq \frac{4 \pi^2}{\lambda^2} \sum_{l=1}^{L} \|\mtx{P}~diag(|\gl|^2)\|_2 \|\Tl u \|^2\\
&= \frac{4 \pi^2}{\lambda^2} \sum_{l=1}^{L} \|\mtx{P}~diag(|\gl|^2)\|_2\|\mathcal{F}\{\Tl \vct{u}\}\|^2\\
&= \frac{4 \pi^2}{\lambda^2} \sum_{l=1}^{L} \|\mtx{P}~diag(|\gl|^2)\|_2 \|\mathcal{F}\{\vct{u}\}_\ell\|^2,\\
\end{align*}
where we first applied Parseval's theorem followed by the Fourier-slice theorem. $\mathcal{F}\{\vct{u}\}_\ell$ denotes the slice in Fourier domain corresponding to angle $\ell$ in spatial domain. To maximize this sum we have to allocate the total energy of $\vct{u}$ at the intersection of all slices, that is $\mathcal{F}\{\vct{u}\}(\vct{k}) = \delta(\vct{k})$. Therefore, the following holds:
\begin{align*}
\sum_{l=1}^{L}\vct{u}^H \jac\gl^H \Big( \sum_{m=1}^{M/L}\vct{a}_m \vct{a}_m^H \Big)\jac\gl \vct{u} \leq \frac{4 \pi^2}{\lambda^2} \sum_{l=1}^{L} \|\mtx{P}~diag(|\gl|^2)\|_2 \|\vct{u}\|^2
\end{align*}
Let $\vct{q}_\ell = -\frac{2 \pi i}{\lambda}\overline{\Tl} \bar{\vct{u}}$, then the second term in Eq. \eqref{qform}  
\begin{align*}
&-\frac{4\pi^2} {\lambda^2}\sum_{l=1}^{L}\Re \Bigg(\vct{u}^H \Tl^Hdiag\Big[\Big( \pdiff{\gl}\mathcal{L}_{\epsilon}(\vct{x}) \Big)^H \odot \cgl \Big]\overline{\Tl}\bar{\vct{u}}\Bigg) \Bigg]
=\sum_{l=1}^{L}\Re \Big( \vct{q}_\ell^T \mtx{D}_\ell \vct{q}_\ell\Big)\\
&\leq \sum_{\ell=1}^{L} \||\mtx{D}_\ell|\|_2 \|\vct{q}_\ell\|^2 =\frac{4 \pi^2}{\lambda^2} \sum_{\ell=1}^{L} \||\mtx{D}_\ell|\|_2~ \|\overline{\Tl} \bar{\vct{u}}\|^2\\
&=\frac{4 \pi^2}{\lambda^2} \sum_{\ell=1}^{L} \||\mtx{D}_\ell|\|_2~ \|\Tl \vct{u}\|^2 
\leq \frac{4 \pi^2}{\lambda^2 } \Big(\sum_{\ell=1}^{L} \||\mtx{D}_\ell|\|_2\Big) \|\vct{u}\|^2 \\
\end{align*}

To summarize the above results, we conclude that
\begin{align*}
&\begin{bmatrix}
\vct{u} \\
\bar{\vct{u}}
\end{bmatrix}^H \nabla^2\mathcal{L}_{\epsilon}(\vct{x})
\begin{bmatrix}
\vct{u} \\
\bar{\vct{u}} 
\end{bmatrix} \leq \\
&\leq 2 \frac{4 \pi^2}{\lambda^2} \Bigg[\sum_{l=1}^{L}  \|\mtx{P}~diag(|\gl|^2)\|_2+   \||\mtx{D}_\ell|\|_2\Bigg]  \|\vct{u}\|^2 \\
&=\frac{4 \pi^2}{\lambda^2} \Bigg[\sum_{l=1}^{L}  \|\mtx{P}~diag(|\gl|^2)\|_2+   \||\mtx{D}_\ell|\|_2\Bigg]
\left\|\begin{bmatrix}
\vct{u} \\
\bar{\vct{u}}
\end{bmatrix}\right\|^2 \\
&=\frac{4 \pi^2}{\lambda^2} \Bigg[\sum_{l=1}^{L}  \left\|\sum_{k=1}^{K} diag(|\vct{p}_k|^2)~diag(|\gl|^2)\right\|_2+   \left\|diag\Big[\Big( \pdiff{\gl}\mathcal{L}_{\epsilon}(\vct{x}) \Big)^H \odot \cgl \Big]\right\|_2\Bigg]
\left\|\begin{bmatrix}
\vct{u} \\
\bar{\vct{u}}
\end{bmatrix}\right\|^2 
\end{align*}
The final result is an iteration-dependant upper bound on the loss Hessian singular value that motivates our practical step size selection in \eqref{eq:Lt}. However, for the following convergence results to hold we need to find an upper bound that is satisfied in each iteration. First, note that $\infnorm{\gl} \leq 1$, reflecting the fact that a passive medium can only attenuate the incident beam. Hence,
\begin{equation*}
\left\|\sum_{k=1}^{K} diag(|\vct{p}_k|^2)~diag(|\gl|^2)\right\|_2 \leq \left\|\sum_{k=1}^{K} diag(|\vct{p}_k|^2)\right\|_2
\end{equation*}
Notice that
\begin{equation*}
\Big( \pdiff{\gl}\mathcal{L}_{\epsilon}(\vct{x}) \Big)^H \odot \cgl = diag(\cgl) \mtx{A}^H(\mtx{A} \gl - \vct{y}_\ell \odot sgn(\mtx{A} \gl)).
\end{equation*}
We are going to bound the $\ell_\infty$ norm of each term of the above quantity:
\begin{equation*}
\infnorm{diag(\cgl) \mtx{A}^H\mtx{A} \gl } \leq \infnorm{\mtx{A}^H\mtx{A} \gl} \infnorm{\cgl}  \leq \lambda_{max}(\mtx{A}^H\mtx{A}) \twonorm{\gl} \leq \lambda_{max}(\mtx{A}^H\mtx{A})\sqrt{P},
\end{equation*}
and
\begin{align*}
\infnorm{diag(\cgl) \mtx{A}^H (\vct{y}_\ell \odot sgn(\mtx{A} \gl)} &\leq \infnorm{\mtx{A}^H( \vct{y}_\ell \odot sgn(\mtx{A} \gl))}, \\
&\leq \twonorm{\mtx{A}^H} \twonorm{ \vct{y}_\ell \odot sgn(\mtx{A} \gl)} 
\leq \sqrt{\lambda_{max}(\mtx{A}^H\mtx{A})} \infnorm{\vct{y}_\ell}.
\end{align*}
Therefore, the Hessian spectral norm is upper bounded by
\begin{equation}
\Gamma = \frac{4\pi^2}{\lambda^2}\left( (1+\sqrt{P}) L \left\|\sum_{k=1}^{K} diag(|\vct{p}_k|^2)\right\|_2 + \sqrt{ \left\|\sum_{k=1}^{K} diag(|\vct{p}_k|^2)\right\|_2} \sum_{l =1}^{L} \twonorm{\vct{y}_\ell}\right) ,
\end{equation}
independent of $\tau$.

Let $\mathcal{L}(\vct{x})_{total}^{\epsilon} = \mathcal{L}_{\epsilon}(\vct{x}) + h(\vct{x})$ the smoothed version of the total loss, where $h(\vct{x})$ is an arbitrary convex scalar function.  Using the Wirtinger derivative version of Taylor's approximation theorem on $\mathcal{L}_{\epsilon}(\vct{x})$, the total loss at consecutive iterations can be written as
\begin{align} \label{eq:lossineq}
\mathcal{L}(\vct{x}_{\tau+1})_{total}^{\epsilon} &= \mathcal{L}(\vct{x_{\tau}})_{total}^{\epsilon} + 
\cplxvct{\nabla\mathcal{L}_{\epsilon}(\vct{x}_\tau)}^H \cplxvct{\iterdiff} \nonumber \\
&+ \frac{1}{2}\cplxvct{\iterdiff}^H \left( \int_0^1  \nabla^2\mathcal{L}_{\epsilon}(\vct{x}_\tau + t (\iterdiff)) dt\right)\cplxvct{\iterdiff} + h(\vct{x}_{\tau+1})\nonumber\\
&\leq \mathcal{L}(\vct{x_{\tau}})_{total}^{\epsilon} + 
\cplxvct{\nabla\mathcal{L}_{\epsilon}(\vct{x}_\tau)}^H \cplxvct{\iterdiff} + \frac{\Gamma}{2}\twonorm{\cplxvct{\iterdiff}}^2 + h(\vct{x}_{\tau+1})\nonumber\\
&=  \mathcal{L}(\vct{x_{\tau}})_{total}^{\epsilon} + \mu \cplxvct{(\iterdiff)/{\mu}+\nabla\mathcal{L}_{\epsilon}(\vct{x}_\tau)}^H\cplxvct{(\iterdiff)/{\mu}} \nonumber\\
&-\mu(1-\frac{\Gamma \mu}{2})\twonorm{\cplxvct{(\iterdiff)/\mu}}^2 + h(\vct{x}_{\tau+1})
\end{align}
By the definition of the proximal operator 
\begin{equation*}
\vct{x}_{\tau+1} =\argmin_{\vct{z} \in \mathbb{C}^N} \frac{1}{2 \mu} \twonorm{\vct{z}-(\vct{x}_{\tau} - \mu \nabla\mathcal{L}_{\epsilon}(\vct{x}_\tau) )}^2 + h(\vct{x}_\tau) = \vct{x}_\tau - \mu \vct{G}_\tau(\vct{x}_\tau),
\end{equation*}
where $\vct{G}_\tau(\vct{x})$ is the generalized gradient at $\vct{x}$ in iteration $\tau$. 
Due to the necessary condition of optimality, we must have 
\begin{equation*}
\vct{z}  - (\vct{x}_{\tau} - \mu \nabla\mathcal{L}_{\epsilon}(\vct{x}_\tau) + t \vct{v} = 0,
\end{equation*}
where $\vct{v} \in \partial h(\vct{z})$ is a subgradient of $h(\vct{x})$ at $\vct{z}$. Substituting $\vct{z} =  \vct{x}_\tau - \mu \vct{G}_\tau(\vct{x}_\tau)$ yields
\begin{equation*}
\vct{v} = \vct{G}_\tau(x_\tau)-\nabla\mathcal{L}_{\epsilon}(\vct{x}_\tau) = -(\iterdiff)/\mu-\nabla\mathcal{L}_{\epsilon}(\vct{x}_\tau)
\end{equation*}
Since $h(\vct{x})$ is convex and $\vct{v} \in \partial h(\vct{x}_{\tau+1})$ we have
\begin{equation*}
h(\vct{x}_{\tau+1}) \leq h(\vct{x}_{\tau}) + \mu \cplxvct{-(\iterdiff)/\mu-\nabla\mathcal{L}_{\epsilon}(\vct{x}_\tau)}^H \cplxvct{(\iterdiff)/\mu}
\end{equation*}
Combining this result with \eqref{eq:lossineq} yields
\begin{align*}
\mathcal{L}_{total}^{\epsilon}(\vct{x}_{\tau +1}) - \mathcal{L}_{total}^{\epsilon}(\vct{x}_\tau)  &\leq  - \mu (1 - \frac{\Gamma  \mu}{2}) \twonorm{\cplxvct{(\iterdiff)/\mu}}^2 \\
&\leq -\frac{\Gamma}{2}\twonorm{\cplxvct{\iterdiff}} ^2 = -\Gamma \twonorm{\iterdiff}^2
\end{align*}
Summing over both sides up to some fixed iteration $T$ we have
\begin{align*}
\Gamma \sum_{\tau =0}^T \twonorm{\iterdiff}^2 \leq \mathcal{L}_{total}^{\epsilon}(\vct{x}_0) -\mathcal{L}_{total}^{\epsilon}(\vct{x}_{T +1}) \leq \mathcal{L}_{total}^{\epsilon}(\vct{x}_0) -\mathcal{L}_{total}^{\epsilon}(\vct{x}^*)
\end{align*}
for a global minimizer $\vct{x}^*$ of $\mathcal{L}_{total}^{\epsilon}(\vct{x})$. Since the above expression holds for any $\epsilon$, we take $\epsilon \rightarrow 0$ and obtain
\begin{equation*}
\Gamma \sum_{\tau =0}^T \twonorm{\iterdiff}^2  \leq \mathcal{L}_{total}(\vct{x}_0) -\mathcal{L}_{total}(\vct{x}^*)
\end{equation*}
Since the series on the left hand side converges, we must have
\begin{equation*}
\lim_{\tau \rightarrow \infty} \twonorm{\vct{x}_{\tau+1}- \vct{x}_{\tau}} = \lim_{\tau \rightarrow \infty} \twonorm{prox_h(\vct{z}_{\tau})- \vct{x}_{\tau}} = 0.
\end{equation*}
Moreover, 
\begin{equation*}
\Gamma \sum_{\tau =0}^T \twonorm{\iterdiff}^2 \geq \Gamma (T+1) \min_{\tau \in \{0,1,..,T\}} \twonorm{\iterdiff}^2 
\end{equation*}
and therefore
\begin{equation*}
 \min_{\tau \in \{1,2,..,T\}} \twonorm{prox_h(\vct{z}_{\tau})- \vct{x}_{\tau}}  \leq \frac{\mathcal{L}_{total}(\vct{x}_0) -\mathcal{L}_{total}(\vct{x}^*)}{\Gamma (T+1)} \leq \frac{\mathcal{L}_{total}(\vct{x}_0) -\mathcal{L}_{total}(\vct{x}^*)}{\mu (T+1)}.
\end{equation*}
We conclude the proof of Theorem \ref{thm:main} by picking $h(\vct{x}) = \textbf{TV}_{3D}(\vct{x}; \vct{w} )$, which is a convex function of $\vct{x}$. Note that the same proof methodology works for any other convex regularizer, and includes total-variation as a special case.
	%

	

\end{document}